\documentclass[12pt,aps,prd,superscriptaddress,showpacs,amsmath,amssymb]{revtex4}
\pdfoutput=1
\usepackage{epsf}
\usepackage{latexsym}
\usepackage{amsmath}
\usepackage{amsfonts}
\usepackage{amssymb}
\usepackage{graphicx}
\def\:{\ \ \ }
\newcommand{\be}{\begin{eqnarray}}
\newcommand{\ee}{\end{eqnarray}}

\usepackage{lscape}
\usepackage{graphics}
\usepackage{epsfig}
\usepackage{graphicx}
\usepackage{rotate}
\usepackage{color}
\usepackage{amsmath, amsthm} 

\def\simle{\lower 2pt \hbox {$\buildrel < \over {\scriptstyle \sim }$}}
\def\simge{\lower 2pt \hbox {$\buildrel > \over {\scriptstyle \sim }$}}

\begin{document}

\title{A COMPREHENSIVE MODEL OF DARK ENERGY, INFLATION AND BLACK HOLES}

\author{PETER L. BIERMANN$^*$ }

\address{Department of Physics and Astronomy, The University
of Alabama, Box 870324, Tuscaloosa, AL 35487-0324, USA\\
 MPI for Radioastronomy, Bonn, Germany\\
Karlsruhe Institute of Technology (KIT) - Institut f{\"u}r Kernphysik, Germany\\
 Department of Physics, University of Alabama at Huntsville, AL, USA\\
Department of Physics \& Astronomy, University of Bonn, Germany\\
$^*$E-mail: plbiermann@mpifr-bonn.mpg.de}

\author{BENJAMIN C. HARMS}

\address{ Department of Physics and Astronomy,
 The University of Alabama,\\
Box 870324, Tuscaloosa, AL 35487-0324, USA\\
E-mail:bharms@bama.ua.edu}	

\begin{abstract}
We derive two new equations of quantum gravity and combine them with reinterpretations of previously proposed concepts of dark energy, black holes, inflation, the arrow of time and the energy at which rest-mass first manifests itself into a theory which may be a first step toward a comprehensive description of all these phenomena.  The resulting theory also predicts new tests which can be experimentally checked within a few years. 
\end{abstract}

\vspace*{10mm}

\maketitle

Our goal is to provide the first step towards a description of a comprehensive model of the birth and evolution of the universe \cite{BH}.  One of the new equations is the creation equation. The other is the Lorentz factor expression for the motion of the soliton shell front of emitted gravitons at merger or birth of a new black hole.

We interpret the usual notion of a quasi static background field as one filled with gravitons.    A possible model which has the features we require is two four-branes embedded in a five-dimensional bulk and separated by the Planck length; such a model is inspired by but different from the Randall-Sundrum models \cite{RS99}.  A metric which describes the transfer of energy from a four-brane where gravity is strong to the weak four-brane on which we live is given by
\begin{eqnarray}
ds^2 = -e^{(u/l)^m\,t/\psi}\,c^2\,dt^2\,+\,{{\rm e}^{ \left( 1-b \left( {\frac {u}{l}} \right) ^{n} \right)\,2\,t/ {
\alpha}{{\it t_H}}}}\,{{\rm e}^{- \left( 
{\frac {u}{l}} \right) ^{k} \left( 1-{\frac {t}{\phi}} \right) }}\,dx_i\,dx^i + {{\rm e}^{ \left( {\frac {u}{l}}
 \right) ^{p}\,t/{\beta}}}\,du^2 \,
\label{metric} 
 \end{eqnarray}
where $i\,=\,1,2,3$, $\tau_H$ is the Hubble time, $l$ is the Planck length, $u$ is the coordinate in the fifth dimension, and the remaining, non-coordinate quantities are arbitrary parameters.  Although the five-dimensional covariant divergence of the energy-momentum tensor does not vanish everywhere, it vanishes on the weak-gravity (weak) brane ($u = 0$), and it is approximately zero on the strong-gravity (strong) brane ($u = l$) for large $\beta/\tau_{H}$ and $\psi/\tau_{H}$.  This metric describes a weak brane which is expanding with time and a strong brane which is contracting with time, albeit very slowly for the latter brane.  The cosmological constant measured on the weak brane is $\Lambda_{w-g}\,=\,-3\,{\frac {1}{{\alpha}^{2}{{\it t_H}}^{2}}}$.

In four space-time dimensions the relativistic Boltzmann equation on the weak brane can be written in terms of a parameter $x= \frac{h\,\nu_0}{k_B\,T_0}$ where $\nu_{0}$ is the frequency of a wave at emission, and $T_0$ is a temperature which is characteristic of the background source of energy.  In terms of $x$ the Boltzmann equation for the number density on the weak brane, ${\mathcal N}(x,t)$,  is
\begin{eqnarray}
\frac{\partial}{\partial t}{\mathcal N}(x,t) = \frac{ \kappa_{0}\,c}{k_B\,T_{0}}\frac{T_{g 0}}{T_0}\frac{1}{x}\frac{\partial}{\partial x}{\mathcal N}(x,t)\, ,
\end{eqnarray}
\noindent where $T_{g 0}$ is the initial graviton temperature, and $\kappa_{0}$ is the initial phase space integral of the matrix element squared for quadrupole emission of a graviton wave from the background. The solution of this equation is
\begin{eqnarray}
{\mathcal N}(x,y)\,=\,\frac{1}{\pi^2} \, \frac{x^{3}}{e^X - 1} 
\end{eqnarray}
with $X \, = \, \sqrt{\{x^2 + 2 y\}}$
and
\begin{eqnarray}
y = \int_0^t\,\frac{\kappa_{0}\,c}{k_B\,T_{0}}\frac{T_{g0}}{T_0}\, dt' 
\label{defy}
\end{eqnarray}
where $y$ is not significant until it is on the order of the dimensionless frequency $x$; from information limit arguments one can show that $y$ corresponds to the Higgs mass, the energy at which rest mass first becomes important.  This implies that in the late universe the spectrum is steepened.  Invoking a resonance at energies below this implies a reduced density, and so a reduced energy transfer.  The creation equation is derived from a phase space resonance analysis of the various quantities contributing to the creation of gravitons. 

In the construction of the creation equation we assume that a local comoving frame exists in which the energy is transferred from the background.  This creation equation describes the transfer of energy from the strong brane to the weak brane by resonance; the Lorentz factors appear as we describe the rate of energy transfer.

The creation equation can be written in the form of energy per unit time created in the form of resonant gravitons
\begin{eqnarray}
\frac{d E}{d \tau} \; = \; {\left( N_{GW}  \, \frac{\sigma_{Pl}}{4 \pi R_{s}^{2}} \, \Gamma_{D} \right)}^{2} \, \left( \frac{4 \pi R_{s}^{2}}{\pi \lambda^{2}} \Gamma_{D} \right) \, E_{GW} \Gamma_{E} \, \tau_{Pl}^{-1}
\label{creationeq}
\end{eqnarray}
in the comoving frame with proper time $\tau$ on a spherical surface of radius $R_s$ with graviton number $N_{GW}$.    The Planck area and time are denoted by $\sigma_{Pl}$ and $\tau_{Pl}$.  $\Gamma _D$ is a Lorentz factor applied to the surface density relative to the observer and to the size of a patch in the comoving frame  $\Gamma_{E}$ is the Lorentz factor as applied to the transformation of energy.  For gravitons $\Gamma_{D} \, = \, \Gamma_{E} $.  $E_{GW}$ is the energy of the gravitons on the surface in the observer frame, and $\lambda \, \sim \, c h/E_{GW}$ is the corresponding length scale.  The  Lorentz factor of the shell front can be derived as well from information limit arguments and is in the frame of the freshly born black hole is ($r(z, z_{\star})$ is the spatial integral from redshift $z$ to $z_{\star}$ and $H(z)$ is the Hubble parameter)
\begin{eqnarray}
\Gamma \; = \;  \frac{1}{2} \, \frac{r(z, z_{\star})}{l_{Pl}} \, \{H(z) \tau_{Pl}\}^{1/2} \, .
\end{eqnarray}
\par 
A soliton shell front is born in the formation of a black hole, or analogously, the merger of two black holes.  
We are assuming that most of the energy released in the merger of two black holes emerges in the last few characteristic collapse times, so that the soliton shell front has a thickness of just a few wavelengths in the observer frame.  Using the observed space density of supermassive black holes today we find a numerical value for dark energy, which is consistent with observation to within the rather large error bars.
\par 
We picture a black hole as a ``Planck shell" full of gravitons at potential depth $M_{BH}/m_{Pl}$ in terms of a Lorentz factor.  This is rather similar to the concept of a ``stretched horizon" \cite{pri,sus}.  This shell is an impenetrable barrier, and the ensemble of these gravitons constitutes the black hole.  Space-time ends at the inner surface of the shell; there is no physical meaning to space-time coordinates in the region enclosed by the shell.    This concept was later also suggested in terms of a ``firewall" \cite{firewall1}.
\par 
The model for graviton creation described above may also explain the inflationary period \cite{guth81,linde812}.  During inflation new gravitons are produced at the horizon in a ``horizon shell", provided that the newly created gravitons lose their energy adiabatically with a time scale proportional to the Planck time, independent of redshift within the inflationary period, which ends when the Higgs boson begins to appear.
\par 
There are three experimental tests which can be performed to determine the validity of these ideas: i)  pulsar timing arrays (PTAs) to detect the gravitational wave background which makes up dark energy, ii) Lunar laser ranging to detect the single solitons coming through, and iii) ultra high precision clocks to detect the noise due to the individual solitons coming through.
\par 
In our model time is driven by the energy transfer from the Planck sea to our universe.  Thus the arrow of time is uni-directional, since in our model energy flows in only one direction - from the background into our universe.  An immediate consistency check of our model is the question of whether or not it describes a universe which is connected to the background Planck sea at all times.
\par 
The Lorentz factor for the effective propagation of a signal derived above should apply to the propagation of any signal carried by mass-less boson particles, or waves travelling with the speed of light even at distances as small as sub-atomic scales.  In our model such signals never travel at exactly the speed of light, but always just very slightly below.

\acknowledgements
This research was supported in part by the DOE under grant DE-FG02-10ER41714.

\end{document}